\documentclass[
    aps,
    prb,
    10pt,
    twocolumn,
    floatfix,
    superscriptaddress,
]{revtex4-2}

\usepackage{amssymb,amsmath}
\usepackage{booktabs}
\usepackage{dcolumn}
\usepackage{glossaries}
\usepackage{graphicx}
\usepackage[version=4]{mhchem}
\usepackage[per-mode=symbol]{siunitx}
\usepackage[hidelinks]{hyperref}

\newacronym{cn}{CN}{coordination number}
\newacronym{dft}{DFT}{density functional theory}
\newacronym{hdp}{HDP}{halide double perovskite}
\newacronym{md}{MD}{molecular dynamics}
\newacronym{mlip}{MLIP}{machine-learned interatomic potential}
\newacronym{nep}{NEP}{neuroevolution potential}
\newacronym{sed}{SED}{spectral energy density}

\newcolumntype{d}{D{.}{.}{-1}}
\DeclareSIUnit\angstrom{\text{Å}}
\DeclareSIUnit\fu{\text{f.u.}}

\makeatletter
\newcommand{\printfnsymbol}[1]{%
  \textsuperscript{\@fnsymbol{2}}%
}

\begin{document}

\title{Octahedral Tilting in Halide Double Perovskites: \texorpdfstring{\\}{}Disentangling Lone-Pair Chemistry and Geometric Effects}

\author{Mehmet Baskurt}
\author{Erik Fransson}
\author{Madeleine Lindvik}
\author{Paul Erhart}
\author{Julia Wiktor}
\email{julia.wiktor@chalmers.se}
\affiliation{Department of Physics, Chalmers University of Technology, SE-412 96 Gothenburg, Sweden} 

\begin{abstract}
Halide double perovskites (HDPs) have emerged as promising alternatives to their lead-based counterparts.
However, their structural dynamics is less explored than that of conventional halide perovskites.
In this work, we investigate octahedral tilting at \qty{0}{\kelvin} and the relative stability of tetragonal and cubic phases of a set of 57 \glspl{hdp}.
By combining structural and energetic descriptors with simple geometric metrics, we identify the main trends controlling the stabilization of one-tilt tetragonal phases across this family.
We find that both the magnitude of the tilt angles and the energetic preference for tilted phases correlate primarily with the Goldschmidt tolerance factor $t$.
The presence of $ns^2$ lone-pair cations also correlates with enhanced tilting; however, this trend largely reflects that lone-pair chemistries in \glspl{hdp} occur together with ionic sizes that shift $t$ away from unity.
Consistent with this picture, we observe several compounds without lone pairs that nonetheless exhibit strong octahedral tilting.
Finally, using machine-learned interatomic potentials, we connect the \qty{0}{\kelvin} tilting energetics to finite-temperature behavior: compounds with more strongly stabilized tilt phases exhibit higher transition temperatures, and phonon spectra at \qty{350}{\kelvin} reveal soft and broad modes that are consistent with the trends in tolerance factors, tilt angles, and tilt energies at \qty{0}{\kelvin}.
Our results provide a systematic reference for structure--stability relationships in \glspl{hdp} and clarify that lone-pair chemistry is correlated with, rather than the primary cause of, octahedral tilting.
\end{abstract}

\maketitle

\section{Introduction}

Halide perovskites, known for their exceptional optoelectronic properties, have attracted significant interest, especially for photovoltaic applications \cite{kovalenko2017properties, chen2015under, lei2021metal, du2021lead, zhang2016metal, jena2019halide}.
However, concerns surrounding the toxicity of lead-based halide perovskites, as well as stability issues, have spurred intense search for alternative compositions \cite{babayigit2016toxicity, lyu2017addressing, slavney2017chemical, park2019intrinsic}.
\Glspl{hdp} have emerged as promising lead-free alternatives \cite{igbari2019progress, chu2019lead, tang2021lead, lei2021lead}.
Compared to lead-based single perovskites, \glspl{hdp} offer several advantages, including reduced toxicity and, in many cases, enhanced structural and environmental stability \cite{volonakis2016lead, meyer2018lead}.
In addition, the chemical flexibility of \glspl{hdp} enables access to a wide range of compositions with tunable electronic structures \cite{slavney2018small, guo2021machine, khalfin2019advances}.

Beyond composition, lattice dynamics play a crucial role in determining the optoelectronic performance of halide perovskites.
Different \glspl{hdp} can exhibit markedly different degrees of dynamical disorder \cite{tailor2022dielectric, maughan2018tolerance}.
While some \glspl{hdp} have been reported to display reduced anharmonicity relative to lead-based halide perovskites \cite{cohen2022diverging}, other compounds show pronounced anharmonic behavior \cite{cappai2024strong}.
Notably, even closely related halide perovskites can exhibit qualitatively different expressions of anharmonicity \cite{cohen2022diverging}.
This motivates a systematic study across the \gls{hdp} chemical space to identify the factors governing lattice softness and structural dynamics. 

Octahedral tilting instabilities play an important role in the structural and electronic properties of \glspl{hdp} \cite{li2017high}.
Although many halide perovskites adopt a cubic phase at high temperatures, the transition from cubic to lower-symmetry phases (often tetragonal) can occur upon cooling.
Such transitions can influence the bandgap, electronic density of states, and carrier mobility, thereby complicating materials design \cite{zhang2019tuning, li2017high}.
Therefore, it is crucial to better understand the structural and dynamical properties of \glspl{hdp}.

\begin{figure}
\centering
\includegraphics[width=8.5cm]{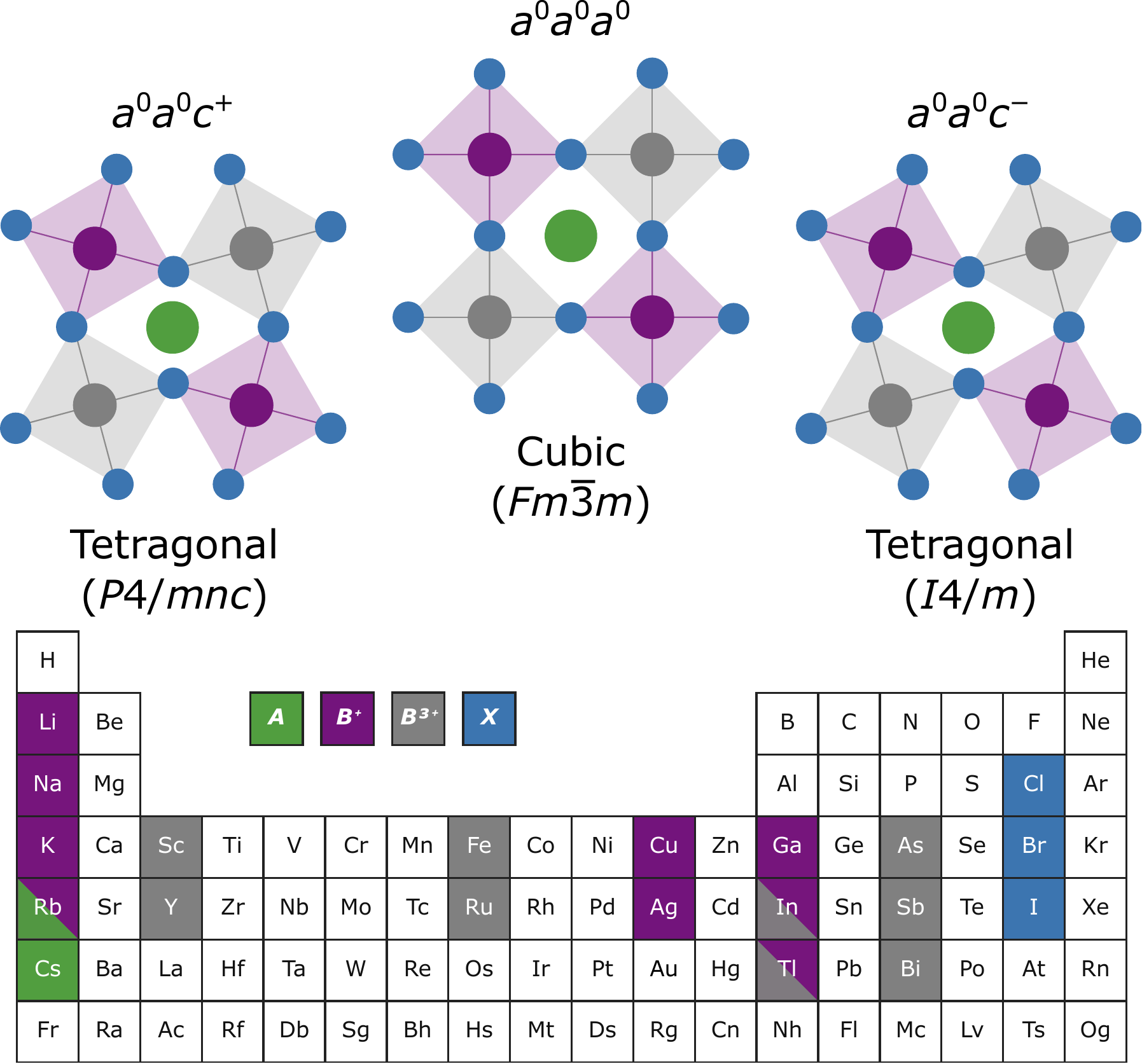}
\caption{
  Overview of the double-halide perovskite structures considered in this work.
  Schematic crystal structures are shown for the tetragonal $P4/mnc$ ($a^0a^0c^+$), cubic $Fm\overline{3}m$ ($a^0a^0a^0$), and tetragonal $I4/m$ ($a^0a^0c^-$) phases.
  The lower panel maps the chemical space explored, highlighting the selected $A$, $B'/B''$, and halide ($X$) components.
}
\label{fig:1}
\end{figure}

Significant efforts have been dedicated to investigating the factors that contribute to the remarkable performance and structural dynamics of halide perovskites.
One factor that has been proposed to contribute to structural softness and instabilities is the electron configuration of the octahedral cations.
In particular, cations with an $ns^2$ electron configuration possess lone-pair electrons that can be stereochemically expressed, and such expression has been discussed as a possible contributor to structural distortions and anharmonicity \cite{fabini2020underappreciated, gao2021metal}.
At the same time, octahedral tilting is also strongly affected by geometric factors such as ionic size and tolerance-factor arguments \cite{lee2016resolving}, and tilting distortions can occur even in compositions without lone-pair cations \cite{maughan2018tolerance, caicedo2024disentangling}.
Disentangling correlation from causation is therefore important when assessing the role of lone-pair chemistry across different materials families.
Here, we employ first-principles \gls{dft} to systematically investigate octahedral tilting and phase stability in a set of 57 \glspl{hdp} spanning a wide range of chemistries and tolerance factors (\autoref{fig:1}).
By optimizing cubic and symmetry-distinct tetragonal cells, we evaluate octahedral tilt angles and the relative stability of the relevant phases.
We then relate these quantities to both the Goldschmidt tolerance factor $t$ and the presence of lone-pair cations.
Furthermore, using \glspl{mlip} based on the \gls{nep} framework, we perform \gls{md} simulations to study phase transitions in representative materials and connect the \qty{0}{\kelvin} tilting energetics to finite-temperature behavior, including phonon \gls{sed} analysis at \qty{350}{\kelvin}.
By using \glspl{hdp} as a chemically diverse yet structurally consistent platform, our work provides a systematic reference for structure--stability relationships and clarifies that lone-pair chemistry is correlated with, rather than required for, octahedral tilting in halide perovskites.

\section{Computational Methodology}
\label{section:computational}

\textbf{\Gls{dft} calculations.}
We performed first-principles calculations within \gls{dft} to investigate the structural properties of \glspl{hdp} using the Vienna \textit{ab initio} Simulation Package \cite{kresse1993ab, kresse1996efficient}. We employed the SCAN meta-GGA functional \cite{sun2015strongly} together with the revised rVV10 nonlocal van der Waals correction (SCAN+rVV10) using the parameterization of Peng \textit{et al.} \cite{versatile2016peng}. We note that the choice of exchange--correlation functional can influence predicted lattice parameters, tilt magnitudes, and relative phase energetics in halide perovskites.\cite{FraWikErh23,wiktor2023quantifying,verdi2023quantum} For halide single perovskites, SCAN+rVV10 has been shown to yield phase transition temperatures in good agreement with experiment \cite{FraWikErh23}. Calculations were carried out for the cubic (F$m\bar{3}m$) and tetragonal ($I$4$/m$, and $P$4$/mnc$) phases of \glspl{hdp}.
For the \qty{0}{\kelvin} structural optimizations and energy differences reported in this work, we used a plane-wave energy cutoff of \qty{720}{\electronvolt} and a $\Gamma$-centered \textit{k}-point mesh generated using a reciprocal-space spacing of \qty{0.15}{\per\angstrom}. Structural optimizations were performed allowing full relaxation of atomic positions and lattice parameters. Additionally, \gls{dft} calculations were performed to generate the input and training dataset for construction of \glspl{mlip}.
For this purpose, we employed a reduced cutoff of \qty{520}{\electronvolt} and a \textit{k}-point spacing of \qty{0.25}{\per\angstrom} to enable efficient sampling of large configuration sets.
In the following, octahedral tilting patterns are described using Glazer’s notation \cite{glazer1972classification, howard1998group}.
The symbol (a$^0$a$^0$a$^0$) represents the cubic structure, characterized by the absence of octahedral rotations.
A pattern of the form (a$^0$a$^0$c$^-$) indicates rotations of the octahedra about the $c$ axis with alternating phase in successive layers, while (a$^0$a$^0$c$^+$) corresponds to rotations with the same phase in successive layers.
Although two-tilt phases such as orthorhombic or monoclinic exist in various halide perovskites, we limit our study to zero- and one-tilt rotation phases to simplify the comparison between different materials.

The tilt energies were calculated from the energy difference between the cubic (F$m\bar{3}m$) and the tetragonal ($I$4$/m$ and $P$4$/mnc$) structures as
\begin{equation}
    \label{eq:phase}
	\Delta E^\text{tilt}_\text{tetragonal} = E_\text{tetragonal} - E_\text{cubic}.
\end{equation}
Here, $E_\text{cubic}$ and $E_\text{tetragonal}$ are the total energies of the optimized cubic (F$m\bar{3}m$) and tetragonal ($I$4$/m$ or $P$4$/mnc$) cells, respectively.

For all considered \glspl{hdp}, we calculated the Goldschmidt tolerance factor \cite{goldschmidt1926gesetze}
\begin{equation}
    t = \frac{r_A + r_X}{\sqrt{2}\,\left(r_B^{\text{eff}} + r_X\right)},
\end{equation}
where $r_A$ and $r_X$ are the ionic radii of the $A$-site cation and halide anion, respectively. 
For double perovskites $A_2B^{\prime}B^{\prime\prime}X_6$, the effective $B$-site radius was taken as the arithmetic average
\begin{equation}
    r_B^{\text{eff}} = \frac{r_{B^{\prime}} + r_{B^{\prime\prime}}}{2},
\end{equation}
with $r_{B^{\prime}}$ and $r_{B^{\prime\prime}}$ being the ionic radii of the two octahedrally coordinated $B$-site cations.
Ionic radii were mostly taken from Shannon’s revised effective ionic radii \cite{shannon1976revised}, using coordination numbers (CNs) corresponding to the perovskite environment (12-fold for $A$, 6-fold for $B^{\prime}/B^{\prime\prime}$, and 6-fold for $X$).
For \ce{In+}, we estimated the radius to be \qty{1.27}{\angstrom} for CN$=6$ by extrapolating the values reported for CN$=9$ and 10 by Baloch \textit{et al.} \cite{baloch2021extending}.

\textbf{Tilt angle analysis.}
To analyze the octahedral tilting, we extracted tilt angles of $B^{\prime}X_6$ and $B^{\prime\prime}X_6$ octahedra using the implemented functions of the \textsc{ovito} package \cite{ovito}.
First, all six $B^{\prime}-X$ and $B^{\prime\prime}-X$ bonds were identified in each octahedron.
Next, using the algorithm described by Larsen \textit{et al.} \cite{larsen2016robust}, $B^{\prime}X_6$ and $B^{\prime\prime}X_6$ octahedra were matched to a simple cubic environment, which would occur in the ideal cubic phase of a perovskite crystal, yielding the scaling and rotation necessary for optimal mapping.
The quaternion form of the rotation was translated into Euler angles using the \textsc{SciPy} package \cite{wiktor2023quantifying,scipy}.
Among the possible settings for the internal rotations, we selected a consistent setting that yields an increasing order of magnitude in accordance with Glazer's principle.
In the following, we report, for each structure, the larger of the two tilt angles obtained for the $B^{\prime}X_6$ and $B^{\prime\prime}X_6$ octahedra.

\textbf{\Gls{mlip} construction.}
We constructed \glspl{mlip} using the fourth-generation \gls{nep} (NEP4) scheme \cite{fan2021neuroevolution, song2024general} as implemented in the \textsc{gpumd} package \cite{fan2022gpumd, xu2025mega}.
\Gls{nep} models were trained for three representative \glspl{hdp}, \ce{Cs2AgAlBr6}, \ce{Cs2AgBiBr6}, and \ce{Cs2InBiBr6}.
Training datasets were generated using an iterative active-learning workflow \cite{FraWikErh23}: candidate configurations were sampled from finite-temperature \gls{md} in relevant phases (including cubic and one-tilt tetragonal variants), and selected structures were labeled by \gls{dft}.
Model uncertainty was estimated using a committee model consisting of one \gls{nep} trained on the full dataset and five additional models trained on independent 80/20 train/validation splits; the spread in predicted energies and forces was used to select new configurations.
For \ce{Cs2InBiBr6}, configurations from the identified monoclinic $P2_1/c$ ground-state phase were also included.
Training was performed with a radial cutoff of \qty{8}{\angstrom} and an angular cutoff of \qty{4}{\angstrom}, for \num{300000} generations.

\textbf{\Gls{md} simulations.}
We performed molecular dynamics simulations using the \textsc{GPUMD} package \cite{fan2022gpumd, xu2025mega}.
Simulations were carried out in the NPT ensemble using the stochastic cell rescaling barostat \cite{bernetti2020pressure}.
Starting from the cubic structure, the temperature was decreased from \qty{500}{\kelvin} to \qty{10}{\kelvin} using a time step of \qty{5}{\femto\second}.
After an initial equilibration of \qty{2.5}{\nano\second} at \qty{500}{\kelvin}, the system was continuously cooled over \qty{120}{\nano\second}.
Simulation cells consisted of \num{40000} atoms, corresponding to a \numproduct{10x10x10} supercell of the conventional cubic unit cell (\num{40} atoms).

\textbf{\Acrfull{sed}.}
To assess the phonon dispersion at finite temperature, we evaluated the \gls{sed} from \gls{md} simulations.
We used supercells consisting of \numproduct{24x24x24} repetitions of the primitive cubic unit cell (10 atoms).
The simulations were first equilibrated for \qty{100}{\pico\second} in the NVT ensemble (using the corresponding lattice parameter obtained from NPT simulations), after which the trajectory was sampled every \qty{25}{\femto\second} in the NVE ensemble.
The \gls{sed} was then computed from the resulting trajectory using \textsc{dynasor}.\cite{dynasor1, dynasor2}

\section{Results and discussion}
\label{sec:results}

To establish structure--stability relationships across the \gls{hdp} family, we investigate the tilt energies, $\Delta E^\text{tilt}_\text{phase}$, and tilt angles, $\theta$, of symmetry-distinct one-tilt tetragonal variants ($I$4$/m$ and $P$4$/mnc$) relative to the cubic phase.
Restricting the analysis to zero- and one-tilt phases enables a consistent comparison across compositions and provides a uniform reference for assessing trends in tilting energetics and geometry.
The tilt energies are calculated with respect to the cubic phase.

First, we generate primitive cells for the cubic (Fm$\bar{3}$m) and two symmetry-distinct tetragonal tilt variants ($I$4$/m$ and $P$4$/mnc$) of \glspl{hdp} $A_2B'B''X_6$.
We considered $A$ = Cs and Rb and $X$ = Cl, Br, I.
The $B'$ site is restricted to chemically plausible monovalent cations (Li, Na, K, Rb, Cu, Ag, In$^{+}$, Tl$^{+}$), while $B''$ includes representative trivalent cations spanning d$^0$/d$^{10}$ and $ns^2$ chemistries (Sc, Y, Fe, Ga, In$^{3+}$, Tl$^{3+}$, As, Sb, Bi).
We select compositions that systematically span a wide range of ionic size mismatch, tolerance factors, and lone-pair chemistry in order to assess trends in octahedral tilting energetics and angles.

Next, we carry out \gls{dft} calculations to optimize the structures of \glspl{hdp} with initial tilts (a$^0$a$^0$a$^0$), (a$^0$a$^0$c$^-$), and (a$^0$a$^0$c$^+$) by relaxing both cell parameters and atomic positions.
We determine the tilt energies and tilt angles as described in \autoref{section:computational}.
\Glspl{hdp} with corresponding tilt energies of the tetragonal $I$4$/m$ and $P$4$/mnc$ phases, $\Delta E^\text{tilt}_{I\text{4}/m}$ and $\Delta E^\text{tilt}_{P\text{4}/mnc}$, respectively, are given in \autoref{table1}.
To aid interpretation, the number of $ns^2$ lone pairs (LP) per formula unit is also included.

\begin{table*}
\centering
\small
\caption{
    Number of $ns^2$ lone pairs (LP), relative energies of the tetragonal phases (per formula unit), octahedral tilt angles, and Goldschmidt tolerance factor $t$ for halide double perovskites.
    Entries are grouped by LP and sorted by decreasing $t$ within each group.
    For structures close to the cubic limit, the energy landscape is shallow and small tilt angles and energies are sensitive to numerical settings.
    These values should therefore be interpreted qualitatively.
}
\label{table1}
\begin{tabular*}{0.95\textwidth}{@{\extracolsep{\fill}}lcddddd}
\toprule
Material & LP
& \multicolumn{1}{c}{$\Delta E_{\text{I4/m}}$}
& \multicolumn{1}{c}{$\Delta E_{\text{P4/mnc}}$}
& \multicolumn{1}{c}{$\theta_{\text{I4/m}}$}
& \multicolumn{1}{c}{$\theta_{\text{P4/mnc}}$}
& \multicolumn{1}{c}{$t$} \\
&
& \multicolumn{1}{c}{(\unit{\milli\electronvolt\per\fu})}
& \multicolumn{1}{c}{(\unit{\milli\electronvolt\per\fu})}
& \multicolumn{1}{c}{(\unit{\degree})}
& \multicolumn{1}{c}{(\unit{\degree})}
& \\
\midrule
\ce{Cs2LiInCl6}   & 0 &   0.08 &   5.19 &  0.20 &  0.00 & 1.01 \\
\ce{Cs2NaFeCl6}   & 0 &   0.23 &  -7.37 &  0.07 &  0.03 & 0.99 \\
\ce{Rb2LiScCl6}   & 0 &   0.30 &   5.58 &  2.67 &  1.62 & 0.97 \\
\ce{Cs2AgAlBr6}   & 0 &  -0.07 &  -3.15 &  0.27 &  0.20 & 0.97 \\
\ce{Rb2LiInCl6}   & 0 &  -6.85 &  -4.87 &  7.37 &  5.37 & 0.96 \\
\ce{Cs2AgFeCl6}   & 0 &  -0.93 &  -6.46 &  0.02 &  0.27 & 0.96 \\
\ce{Cs2NaInCl6}   & 0 &   0.06 &  -3.19 &  0.03 &  0.07 & 0.96 \\
\ce{Cs2AgGaBr6}   & 0 &  -0.09 &   1.69 &  0.08 &  0.08 & 0.95 \\
\ce{Cs2AgScCl6}   & 0 &   0.17 &   2.89 &  0.14 &  0.01 & 0.95 \\
\ce{Cs2NaInBr6}   & 0 &  -0.01 &  -1.95 &  1.37 &  1.36 & 0.95 \\
\ce{Cs2NaTlCl6}   & 0 &   0.57 &   3.71 &  1.60 &  0.80 & 0.94 \\
\ce{Cs2AgInCl6}   & 0 &  -0.05 &   0.65 &  0.01 &  0.00 & 0.94 \\
\ce{Cs2AgScBr6}   & 0 &  -0.15 &   1.54 &  0.40 &  0.32 & 0.93 \\
\ce{Cs2AgInBr6}   & 0 &  -1.69 &   1.70 &  1.55 &  1.39 & 0.93 \\
\ce{Cs2AgYCl6}    & 0 &   2.92 &   2.41 &  3.49 &  2.31 & 0.92 \\
\ce{Rb2NaInCl6}   & 0 & -42.76 & -41.72 & 11.43 & 10.94 & 0.92 \\
\ce{Cs2AgTlBr6}   & 0 &   1.05 &   2.91 &  1.15 &  0.20 & 0.91 \\
\ce{Rb2NaInBr6}   & 0 & -129.70 & -121.21 & 13.45 & 13.45 & 0.91 \\
\ce{Cs2KInCl6}    & 0 & -20.58 & -12.84 &  9.93 &  8.91 & 0.90 \\
\ce{Cs2AgYI6}     & 0 &   2.63 &   2.44 &  3.37 &  2.30 & 0.89 \\
\midrule
\ce{Cs2NaSbCl6}   & 1 &  -1.29 &   1.45 &  1.34 &  1.56 & 0.97 \\
\ce{Cs2LiBiCl6}   & 1 &   0.02 &   0.75 &  0.03 &  0.01 & 0.96 \\
\ce{Cs2LiBiBr6}   & 1 &   2.39 &  -1.54 &  4.35 &  1.43 & 0.95 \\
\ce{Cs2CuBiBr6}   & 1 &  -0.79 &  -1.71 &  0.96 &  0.65 & 0.95 \\
\ce{Cs2AgSbCl6}   & 1 &  -2.73 &  -2.88 &  1.39 &  1.56 & 0.94 \\
\ce{Cs2AgSbBr6}   & 1 &   0.76 &   1.33 &  1.90 &  2.67 & 0.93 \\
\ce{Cs2NaBiCl6}   & 1 &   3.92 &   4.41 &  6.33 &  1.63 & 0.92 \\
\ce{Cs2InScBr6}   & 1 & -35.10 & -27.99 & 11.93 & 11.89 & 0.92 \\
\ce{Cs2NaBiBr6}   & 1 & -27.10 & -21.48 &  9.95 &  8.50 & 0.91 \\
\ce{Cs2AgBiCl6}   & 1 &   0.03 &   0.07 &  4.29 &  1.60 & 0.90 \\
\ce{Cs2NaBiI6}    & 1 & -117.54 & -111.60 & 13.36 & 13.42 & 0.89 \\
\ce{Cs2AgBiBr6}   & 1 &  -2.43 &  -0.65 &  7.28 &  5.47 & 0.89 \\
\ce{Cs2AgBiI6}    & 1 & -40.13 & -30.79 & 11.44 & 10.95 & 0.88 \\
\ce{Rb2NaBiCl6}   & 1 & -187.49 & -183.23 & 14.52 & 14.47 & 0.88 \\
\ce{Rb2NaBiBr6}   & 1 & -250.49 & -241.73 & 15.55 & 15.36 & 0.87 \\
\ce{Cs2KBiCl6}    & 1 & -147.24 & -137.34 & 15.13 & 14.23 & 0.87 \\
\ce{Rb2AgBiCl6}   & 1 & -171.96 & -165.38 & 14.60 & 14.47 & 0.86 \\
\ce{Rb2NaBiI6}    & 1 & -355.48 & -352.12 & 16.03 & 16.13 & 0.86 \\
\ce{Cs2KBiBr6}    & 1 & -188.13 & -179.36 & 16.14 & 16.26 & 0.86 \\
\ce{Rb2AgBiBr6}   & 1 & -204.13 & -195.04 & 15.16 & 15.07 & 0.85 \\
\ce{Cs2KBiI6}     & 1 & -265.28 & -258.33 & 16.95 & 16.82 & 0.85 \\
\ce{Cs2RbBiCl6}   & 1 & -274.39 & -267.12 & 19.01 & 18.89 & 0.85 \\
\ce{Rb2AgBiI6}    & 1 & -263.87 & -256.04 & 15.42 & 15.38 & 0.84 \\
\midrule
\ce{Cs2TlAsBr6}   & 2 &   1.21 &   2.75 &  7.82 &  7.55 & 0.91 \\
\ce{Cs2InSbBr6}   & 2 &  -7.37 &  -3.07 &  8.85 &  8.04 & 0.91 \\
\ce{Cs2TlAsI6}    & 2 & -15.73 & -12.92 &  9.44 &  9.25 & 0.89 \\
\ce{Cs2TlSbBr6}   & 2 & -87.44 & -81.01 & 14.27 & 14.13 & 0.88 \\
\ce{Cs2InBiCl6}   & 2 & -76.19 & -64.89 & 14.08 & 13.61 & 0.88 \\
\ce{Rb2InSbCl6}   & 2 & -241.15 & -233.09 & 17.81 & 17.83 & 0.88 \\
\ce{Rb2TlAsBr6}   & 2 & -176.37 & -172.43 & 16.72 & 16.58 & 0.87 \\
\ce{Cs2TlSbI6}    & 2 & -98.53 & -96.47 & 14.25 & 14.11 & 0.87 \\
\ce{Cs2InBiBr6}   & 2 & -54.97 & -50.79 & 12.83 & 12.85 & 0.87 \\
\ce{Rb2InSbBr6}   & 2 & -185.63 & -182.13 & 16.21 & 16.53 & 0.87 \\
\ce{Cs2TlBiCl6}   & 2 & -148.13 & -138.59 & 16.36 & 16.58 & 0.85 \\
\ce{Cs2TlBiBr6}   & 2 & -145.46 & -140.02 & 15.27 & 15.46 & 0.84 \\
\ce{Rb2InBiCl6}   & 2 & -333.65 & -334.60 & 18.97 & 18.37 & 0.84 \\
\ce{Cs2TlBiI6}    & 2 & -158.16 & -154.37 & 15.21 & 15.39 & 0.83 \\
\bottomrule
\end{tabular*}
\end{table*}

\begin{figure}
\centering
\includegraphics[width=7.5cm]{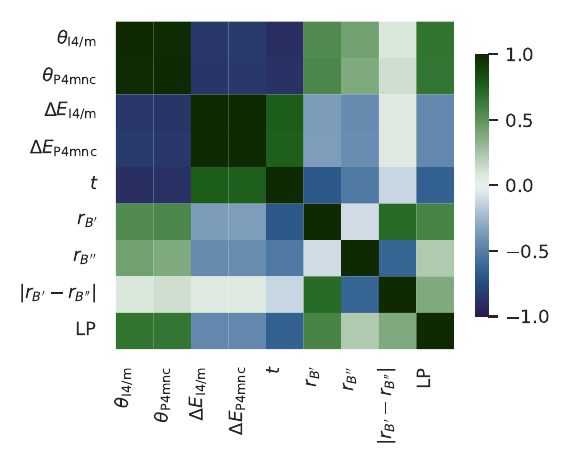}
\caption{
    Pearson correlation matrix ($r$) for computed tilt angles and relative energies of the $I4/m$ and $P4/mnc$ phases, together with selected geometric descriptors ($t$, $r_{B'}$, $r_{B''}$, and $|r_{B'}-r_{B''}|$) and the number of lone pairs (LP).
    Colors indicate $r$ ($-1 \le r \le 1$).
}
\label{fig:corr}
\end{figure}

We classify \glspl{hdp} based on the number of $ns^2$ lone-pair cations on the $B^{\prime}$ and $B^{\prime\prime}$ sites.
Compounds where neither $B^{\prime}$ nor $B^{\prime\prime}$ hosts an $ns^2$ cation are assigned LP = 0.
If exactly one of the $B^{\prime}$ or $B^{\prime\prime}$ cations has an $ns^2$ configuration (e.g., Bi$^{3+}$, Sb$^{3+}$, or As$^{3+}$ on $B^{\prime\prime}$), the compound is assigned LP = 1.
If both $B^{\prime}$ and $B^{\prime\prime}$ host $ns^2$ cations (e.g., \ce{In+} or \ce{Tl+} on $B^{\prime}$ together with Bi$^{3+}$/Sb$^{3+}$/As$^{3+}$ on $B^{\prime\prime}$), the compound is assigned LP = 2.

To identify the key descriptors controlling octahedral tilting across the \gls{hdp} dataset, we first examine correlations between the computed tilt angles and tilt energies and a set of chemical and geometric metrics.
\autoref{fig:corr} shows the Pearson correlation matrix for the full dataset.
A clear correlation is observed between the Goldschmidt tolerance factor $t$ and both the magnitude of the tilt angles and the stabilization energy of the tilted phases: compounds with smaller $t$ generally exhibit larger tilt angles and a stronger energetic preference for tilted structures.
Related tolerance-factor trends have also emerged from previous studies on single perovskites  \cite{yang2017spontaneous,gissler2025influence}.
In contrast, the number of $ns^2$ lone pairs shows a weaker direct correlation with tilting and energetics.
However, LP is itself correlated with ionic-size descriptors and $t$ across the present chemical space, indicating that the apparent LP--tilting trend largely reflects systematic compositional differences in ionic radii rather than an independent driving mechanism.

\begin{figure*}
\centering
\includegraphics[width=\linewidth]{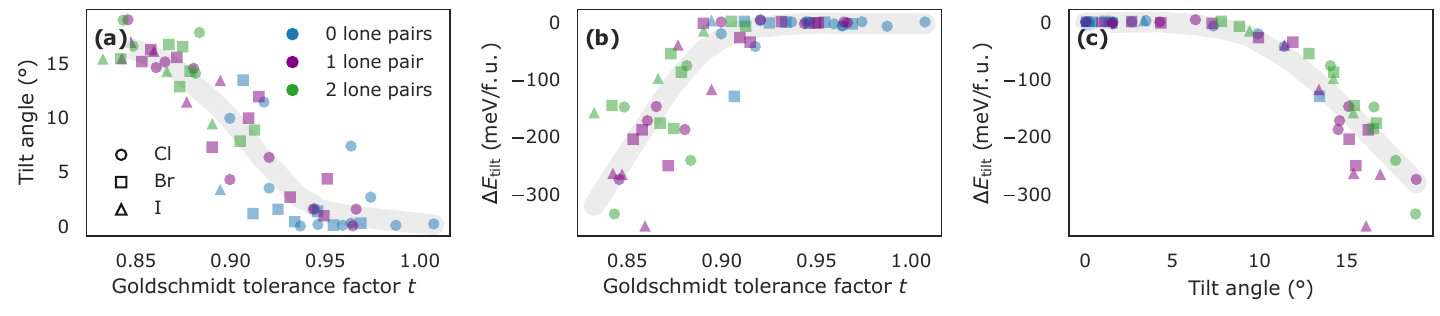}
\caption{
    Relationships between structural distortion and energetic stability in the double-halide perovskite dataset: (a) tilt angle vs.\ Goldschmidt tolerance factor $t$, (b) $\Delta E_{\text{tilt}}$ vs.\ $t$, and (c) $\Delta E_{\text{tilt}}$ vs.\ tilt angle.
    Colors indicate the number of lone pairs (LP) and symbols denote the halide.
    The gray lines serve as a guide to the eye, indicating the overall trend.
}
\label{fig:scatter}
\end{figure*}

These trends are further illustrated in \autoref{fig:scatter}.
Here, we define the tilt energy $\Delta E_{\text{tilt}} = \min(\Delta E_{I4/m}, \Delta E_{P4/mnc})$ as the stabilization energy of the lowest-energy one-tilt tetragonal phase relative to the cubic phase.
In \autoref{fig:scatter}a, the tilt angle increases systematically as $t$ decreases, with only modest scatter between halides and lone-pair classes.
\autoref{fig:scatter}b shows that $\Delta E_{\text{tilt}}$ becomes increasingly negative as $t$ decreases, demonstrating that geometric mismatch provides a strong predictor for the stabilization of octahedral tilting.
Finally, \autoref{fig:scatter}c shows a direct relationship between $\Delta E_{\text{tilt}}$ and tilt angle: larger tilts are generally associated with stronger energetic stabilization of the distorted phase.

\begin{figure*}
\centering
\includegraphics[width=\linewidth]{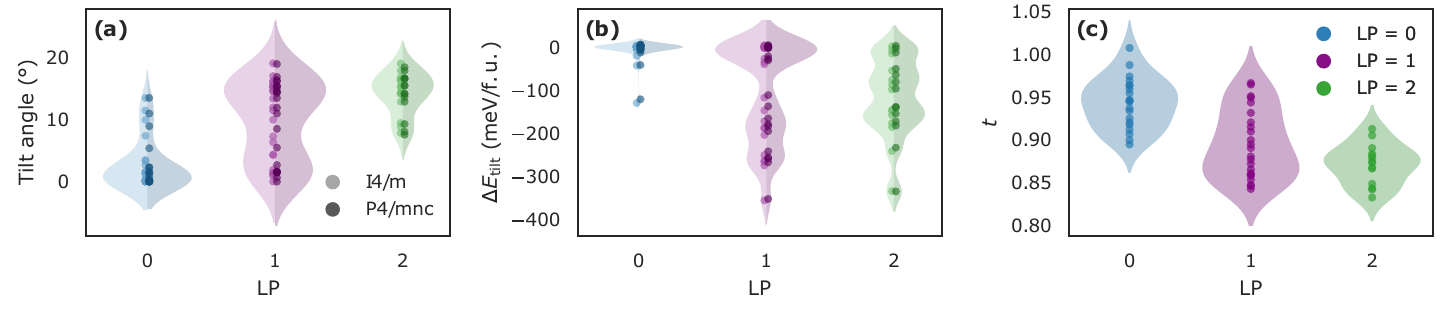}
\caption{
    Distributions of structural, energetic, and geometric descriptors grouped by the number of $ns^2$ lone pairs (LP).
    Violin plots show (a) octahedral tilt angles, (b) relative energies $\Delta E_{\text{tilt}}$, and (c) tolerance factors $t$ for compounds with LP = 0, 1, and 2.
    Points indicate individual compounds.
}
\label{fig:violin}
\end{figure*}

To visualize these trends in a compact form, \autoref{fig:violin} shows distributions of tilt angles, tilt energies, and $t$ grouped by LP.
While compounds with LP = 2 tend to display larger tilt angles and stronger stabilization of the tilted phases than compounds with LP = 0, \autoref{fig:violin}c demonstrates that these groups also occupy different tolerance factor ranges.
This indicates that the LP dependence is, to a significant extent, mediated by systematic differences in ionic radii that shift the tolerance factor away from the cubic ideal.
Conversely, the LP = 0 group contains both nearly cubic materials (with $t \approx 1$) and strongly tilted compounds where $t$ significantly deviates from unity, consistent with a primarily geometric origin of the tilting instability.

\begin{figure}
\centering
\includegraphics[width=\columnwidth]{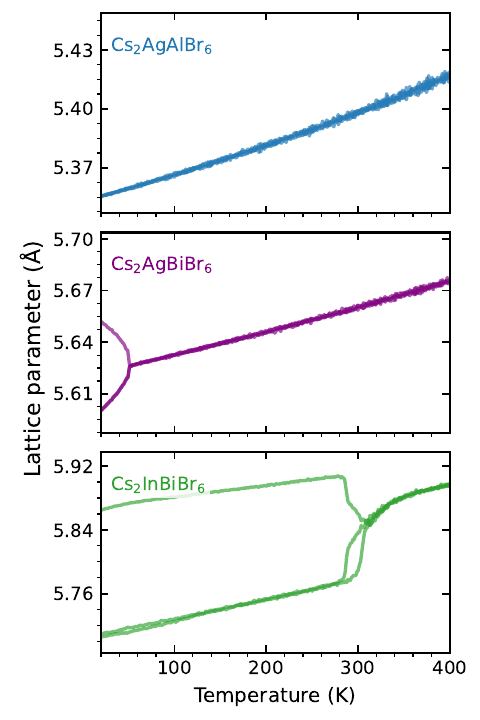}
\caption{
    Temperature dependence of the lattice parameter for representative double-halide perovskites (\ce{Cs2AgAlBr6}, \ce{Cs2AgBiBr6}, and \ce{Cs2InBiBr6}) obtained from cooling simulations. Discontinuities indicate structural phase transitions.
}
\label{fig:lattice}
\end{figure}

\begin{figure*}
\centering
\includegraphics[width=\linewidth]{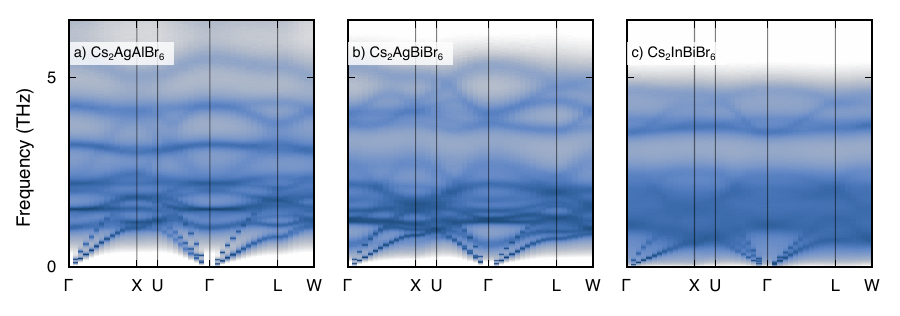}
\caption{
    Phonon \acrfull{sed} at \qty{350}{\kelvin} for the cubic phase of (a) \ce{Cs2AgAlBr6}, (b) \ce{Cs2AgBiBr6}, and (c) \ce{Cs2InBiBr6} along high-symmetry directions in the Brillouin zone.
}
\label{fig:sed}
\end{figure*}

Several compositions illustrate this point.
For example, \ce{Cs2KInCl6}, \ce{Rb2NaInCl6}, and \ce{Rb2NaInBr6} contain alkali metals on the $B^{\prime}$ site and exhibit pronounced tilting despite the absence of lone-pair cations.
These compounds are characterized by tolerance factors substantially below unity, consistent with tilting driven by geometric mismatch and electrostatic packing effects.
Within the LP = 1 group, a wide range of tilt angles and energies is observed, which again tracks variations in $t$ rather than LP alone: materials such as \ce{Cs2AgBiI6} and \ce{Cs2KBiBr6} display large tilts and strongly negative $\Delta E_{\text{tilt}}$, while others such as \ce{Cs2AgBiBr6} exhibit comparatively weak stabilization of the tilted phase.

So far, we have only considered the relaxed structure of \glspl{hdp} at \qty{0}{\kelvin}.
In lead- and tin-based halide perovskites, it has been shown that the degree of octahedral tilting at \qty{0}{\kelvin} correlates with finite-temperature phase transition behavior \cite{wiktor2023quantifying}.
It is therefore useful to connect the \qty{0}{\kelvin} tilting descriptors discussed above (tilt angles and tilt energies) to transition temperatures in representative \glspl{hdp}.
For this purpose, we select three representative compounds spanning the range of tilting tendencies in our dataset: \ce{Cs2AgAlBr6}, \ce{Cs2AgBiBr6}, and \ce{Cs2InBiBr6}.
These materials also span different lone-pair counts (LP = 0, 1, and 2), but, importantly, they differ systematically in their tolerance factors and \qty{0}{\kelvin} tilting energetics.
We perform \gls{mlip}-driven cooling simulations in the NPT ensemble using large supercells containing \num{40000} atoms.
The evolution of the pseudo-cubic lattice parameters with temperature is shown in \autoref{fig:lattice}. 

For \ce{Cs2AgAlBr6}, which is close to the cubic limit with negligible \qty{0}{\kelvin} tilting, the lattice parameter varies smoothly over the full temperature range and no clear structural transition is observed.
In contrast, \ce{Cs2AgBiBr6} and \ce{Cs2InBiBr6}, which exhibit larger tilt angles and more negative tilt energies at \qty{0}{\kelvin}, show clear abrupt changes in the temperature dependence of the lattice parameter, indicative of structural phase transitions.
In \ce{Cs2AgBiBr6}, a tetragonal-to-cubic transition is observed at about \qty{50}{\kelvin} and in \ce{Cs2InBiBr6} it occurs at about \qty{310}{\kelvin}.
These observations support the general trend that stronger stabilization of tilted phases at \qty{0}{\kelvin} (more negative $\Delta E_{\text{tilt}}$ and larger $\theta$) is associated with higher transition temperatures, consistent with a primarily geometric control of the tilting instability across the \gls{hdp} family. 

Beyond the average structural signatures in the lattice parameters, the finite-temperature vibrational spectra provide a complementary view of the lattice softness and anharmonicity.
\autoref{fig:sed} shows the phonon \gls{sed} at \qty{350}{\kelvin} for \ce{Cs2AgAlBr6}, \ce{Cs2AgBiBr6}, and \ce{Cs2InBiBr6} along high-symmetry directions. The plotted path includes the Brillouin-zone boundary points associated with octahedral tilting instabilities in the cubic double-perovskite structure \cite{klarbring2020anharmonicity}.
For \ce{Cs2AgAlBr6}, which remains close to the cubic limit with $t\approx 1$ and negligible \qty{0}{\kelvin} tilting, the \gls{sed} exhibits relatively sharp dispersive features.
In contrast, for \ce{Cs2AgBiBr6} and especially \ce{Cs2InBiBr6}, which display larger tilt angles and stronger stabilization of tilted phases at \qty{0}{\kelvin} (smaller $t$), the low-frequency features across the Brillouin zone become markedly broader and more diffuse.
Such spectral broadening is consistent with increased anharmonicity and reduced phonon lifetimes, indicating a softer lattice and stronger anharmonicity.

Lattice softness and anharmonic lattice fluctuations have been widely discussed as contributing to the favorable optoelectronic properties and defect tolerance of lead-halide perovskites, for example through enhanced dielectric screening, polaron formation, and suppression of nonradiative recombination.\cite{miyata2017lead,chu2020low} In this context, our systematic evaluation of tilt angles and tilt energies provides a reference for comparing the relative strength of octahedral-tilting instabilities and associated trends in lattice softness across HDPs. While the quantitative impact on device performance will depend on detailed band structure and defect physics, the present dataset offers a framework for guiding materials selection. Similar geometric control of tilting via ionic-size mismatch is expected to apply in similar material families, for example oxide and chalcogenide perovskites.

\section{Conclusions}\label{sec:conclusions}

In this study, we have investigated octahedral tilting instabilities and the stability of one-tilt tetragonal phases in a chemically diverse set of halide double perovskites.
By combining \gls{dft} calculations for the cubic and two symmetry-distinct tetragonal variants with statistical analysis of structural and chemical descriptors, we identify robust trends linking tilt angles and tilt stabilization energies to simple geometric metrics.

Across the explored chemical space, the Goldschmidt tolerance factor $t$ emerges as the strongest predictor of both the magnitude of octahedral tilting and the energetic preference for tilted phases.
Compounds with $t$ close to unity generally remain nearly cubic with small tilt angles and weak stabilization of the tetragonal variants, whereas decreasing $t$ is associated with larger tilt angles and increasingly negative tilt energies.
While materials containing $ns^2$ lone-pair cations tend to show stronger tilting on average, we find that this trend is largely mediated by systematic differences in ionic radii: lone-pair chemistry correlates with tilting primarily through its correlation with $t$ within the present dataset.
Importantly, pronounced tilting is also found in several LP = 0 compounds containing alkali metals on the $B^{\prime}$ site, consistent with tilting driven by geometric packing and electrostatic considerations rather than lone-pair activity alone.

To connect the \qty{0}{\kelvin} descriptors to finite-temperature behavior, we trained \glspl{mlip} and performed large-scale NPT cooling simulations for three representative compounds, \ce{Cs2AgAlBr6}, \ce{Cs2AgBiBr6}, and \ce{Cs2InBiBr6}.
The temperature evolution of the lattice parameters reveals clear phase-transition signatures for the more strongly tilted compounds, while \ce{Cs2AgAlBr6} remains close to the cubic limit over the investigated temperature range.
Furthermore, \gls{sed} analysis at \qty{350}{\kelvin} shows progressively broader low-frequency features as $t$ decreases, indicating increased anharmonicity and shorter phonon lifetimes.
Together, these results demonstrate that tilt angles and tilt energies at \qty{0}{\kelvin} provide a useful qualitative indicator of finite-temperature lattice softness and transition behavior, with geometric packing, captured by the tolerance factor, serving as the dominant control parameter across the halide double perovskite family.
\section{Data availability} 
DFT input files, structures, \gls{nep} models, as well as the \gls{dft} databases used to train these models, are available on Zenodo at \url{https://doi.org/10.5281/zenodo.18841767}.

\section{Acknowledgments}

The authors acknowledge funding from the Swedish Strategic Research Foundation through a Future Research Leader programme (FFL21-0129), the Swedish Research Council (Nos.~2019-03993, 2020-04935, and 2025-03999), European Research Council (ERC Starting Grant no. 101162195),
and the Knut and Alice Wallenberg Foundation (Nos.~2023.0032 and 2024.0042).
The computations were enabled by resources provided by the National Academic Infrastructure for Supercomputing in Sweden (NAISS) at PDC, C3SE, and NSC, partially funded by the Swedish Research Council through grant agreement no. 2022-06725.

\bibliography{references}

@article{lee2016resolving,
  title={Resolving the physical origin of octahedral tilting in halide perovskites},
  author={Lee, Jung-Hoon and Bristowe, Nicholas C and Lee, June Ho and Lee, Sung-Hoon and Bristowe, Paul D and Cheetham, Anthony K and Jang, Hyun Myung},
  journal={Chemistry of Materials},
  volume={28},
  number={12},
  pages={4259--4266},
  year={2016},
  publisher={ACS Publications}
}

@article{chu2020low,
  title={Low-frequency lattice phonons in halide perovskites explain high defect tolerance toward electron-hole recombination},
  author={Chu, Weibin and Zheng, Qijing and Prezhdo, Oleg V and Zhao, Jin and Saidi, Wissam A},
  journal={Science advances},
  volume={6},
  number={7},
  pages={eaaw7453},
  year={2020},
  publisher={American Association for the Advancement of Science}
}

@article{miyata2017lead,
  title={Lead halide perovskites: Crystal-liquid duality, phonon glass electron crystals, and large polaron formation},
  author={Miyata, Kiyoshi and Atallah, Timothy L and Zhu, X-Y},
  journal={Science Advances},
  volume={3},
  number={10},
  pages={e1701469},
  year={2017},
  publisher={American Association for the Advancement of Science}
}

@article{klarbring2020anharmonicity,
  title={Anharmonicity and ultralow thermal conductivity in lead-free halide double perovskites},
  author={Klarbring, Johan and Hellman, Olle and Abrikosov, Igor A and Simak, Sergei I},
  journal={Physical review letters},
  volume={125},
  number={4},
  pages={045701},
  year={2020},
  publisher={APS}
}

@article{yang2017spontaneous,
  title={Spontaneous octahedral tilting in the cubic inorganic cesium halide perovskites CsSnX3 and CsPbX3 (X= F, Cl, Br, I)},
  author={Yang, Ruo Xi and Skelton, Jonathan M and Da Silva, E Lora and Frost, Jarvist M and Walsh, Aron},
  journal={The journal of physical chemistry letters},
  volume={8},
  number={19},
  pages={4720--4726},
  year={2017},
  publisher={ACS Publications}
}

@article{verdi2023quantum,
  title={Quantum paraelectricity and structural phase transitions in strontium titanate beyond density functional theory},
  author={Verdi, Carla and Ranalli, Luigi and Franchini, Cesare and Kresse, Georg},
  journal={Physical Review Materials},
  volume={7},
  number={3},
  pages={L030801},
  year={2023},
  publisher={APS}
}

@article{dynasor2,
title = {Dynasor 2: From simulation to experiment through correlation functions},
journal = {Computer Physics Communications},
volume = {316},
pages = {109759},
year = {2025},
issn = {0010-4655},
doi = {https://doi.org/10.1016/j.cpc.2025.109759},
url = {https://www.sciencedirect.com/science/article/pii/S0010465525002619},
author = {Esmée Berger and Erik Fransson and Fredrik Eriksson and Eric Lindgren and Göran Wahnström and Thomas Holm Rod and Paul Erhart},
}

@article{gissler2025influence,
  title={Influence of the Chemical Composition on the Perovskites Anharmonicity: Toward a Stable Inorganic Perovskite},
  author={Gissler, Antoine and Schulz, Philip and Baranek, Philippe},
  journal={Solar RRL},
  pages={e202500690},
  year={2025},
  doi={https://doi.org/10.1002/solr.202500690},
}

@article{shannon1976revised,
  title={Revised effective ionic radii and systematic studies of interatomic distances in halides and chalcogenides},
  author={Shannon, Robert D},
  journal={Foundations of Crystallography},
  volume={32},
  number={5},
  pages={751--767},
  year={1976},
  publisher={International Union of Crystallography}
}

@article{baloch2021extending,
  title={Extending {Shannon}'s ionic radii database using machine learning},
  author={Baloch, Ahmer AB and Alqahtani, Saad M and Mumtaz, Faisal and Muqaibel, Ali H and Rashkeev, Sergey N and Alharbi, Fahhad H},
  journal={Physical Review Materials},
  volume={5},
  number={4},
  pages={043804},
  year={2021},
  publisher={APS}
}

@article{goldschmidt1926gesetze,
  title={{Die Gesetze der Krystallochemie}},
  author={Goldschmidt, Victor Moritz},
  journal={Naturwissenschaften},
  volume={14},
  number={21},
  pages={477--485},
  year={1926},
  publisher={Springer}
}

@article{gao2021metal,
  title={Metal cation s lone-pairs increase octahedral tilting instabilities in halide perovskites},
  author={Gao, Lingyuan and Yadgarov, Lena and Sharma, Rituraj and Korobko, Roman and McCall, Kyle M and Fabini, Douglas H and Stoumpos, Constantinos C and Kanatzidis, Mercouri G and Rappe, Andrew M and Yaffe, Omer},
  journal={Materials Advances},
  volume={2},
  number={14},
  pages={4610--4616},
  year={2021},
  doi = {10.1039/D1MA00288K},
}

@article{dynasor1,
author = {Fransson, Erik and Slabanja, Mattias and Erhart, Paul and Wahnström, Göran},
title = {dynasor—A Tool for Extracting Dynamical Structure Factors and Current Correlation Functions from Molecular Dynamics Simulations},
journal = {Advanced Theory and Simulations},
volume = {4},
number = {2},
pages = {2000240},
doi = {https://doi.org/10.1002/adts.202000240},
url = {https://advanced.onlinelibrary.wiley.com/doi/abs/10.1002/adts.202000240},
year = {2021}
}

@article{kresse1993ab,
  title={Ab initio molecular dynamics for liquid metals},
  author={Kresse, Georg and Hafner, J{\"u}rgen},
  journal={Physical Review B},
  volume={47},
  number={1},
  pages={558},
  year={1993},
  publisher={APS}
}

@article{kresse1996efficient,
  title={Efficient iterative schemes for ab initio total-energy calculations using a plane-wave basis set},
  author={Kresse, Georg and Furthm{\"u}ller, J{\"u}rgen},
  journal={Physical Review B},
  volume={54},
  number={16},
  pages={11169},
  year={1996},
  publisher={APS}
}

@article{sun2015strongly,
  title={Strongly constrained and appropriately normed semilocal density functional},
  author={Sun, Jianwei and Ruzsinszky, Adrienn and Perdew, John P},
  journal={Physical Review Letters},
  volume={115},
  number={3},
  pages={036402},
  year={2015},
  publisher={APS}
}

@article{cohen2022diverging,
  title={Diverging expressions of anharmonicity in halide perovskites},
  author={Cohen, Adi and Brenner, Thomas M and Klarbring, Johan and Sharma, Rituraj and Fabini, Douglas H and Korobko, Roman and Nayak, Pabitra K and Hellman, Olle and Yaffe, Omer},
  journal={Advanced Materials},
  volume={34},
  number={14},
  pages={2107932},
  year={2022},
  publisher={Wiley Online Library}
}

@article{cappai2024strong,
  title={Strong Anharmonicity at the Origin of Anomalous Thermal Conductivity in Double Perovskite \ce{Cs2NaYbCl6}},
  author={Cappai, Antonio and Melis, Claudio and Marongiu, Daniela and Quochi, Francesco and Saba, Michele and Congiu, Francesco and He, Yihui and Slade, Tyler J and Kanatzidis, Mercouri G and Colombo, Luciano},
  journal={Advanced Science},
  volume={11},
  number={9},
  pages={2305861},
  year={2024},
  publisher={Wiley Online Library}
}

@article{versatile2016peng,
  title = {Versatile van der Waals Density Functional Based on a Meta-Generalized Gradient Approximation},
  author = {Peng, Haowei and Yang, Zeng-Hui and Perdew, John P. and Sun, Jianwei},
  journal = {Physical Review X},
  volume = {6},
  issue = {4},
  pages = {041005},
  numpages = {15},
  year = {2016},
  month = {Oct},
  publisher = {American Physical Society},
  doi = {10.1103/PhysRevX.6.041005},
  url = {https://link.aps.org/doi/10.1103/PhysRevX.6.041005}
}

@article{kovalenko2017properties,
  title={Properties and potential optoelectronic applications of lead halide perovskite nanocrystals},
  author={Kovalenko, Maksym V and Protesescu, Loredana and Bodnarchuk, Maryna I},
  journal={Science},
  volume={358},
  number={6364},
  pages={745--750},
  year={2017},
  publisher={American Association for the Advancement of Science}
}

@article{chen2015under,
  title={Under the spotlight: The organic--inorganic hybrid halide perovskite for optoelectronic applications},
  author={Chen, Qi and De Marco, Nicholas and Yang, Yang Michael and Song, Tze-Bin and Chen, Chun-Chao and Zhao, Hongxiang and Hong, Ziruo and Zhou, Huanping and Yang, Yang},
  journal={Nano Today},
  volume={10},
  number={3},
  pages={355--396},
  year={2015},
  publisher={Elsevier}
}

@article{lei2021metal,
  title={Metal halide perovskites for laser applications},
  author={Lei, Lei and Dong, Qi and Gundogdu, Kenan and So, Franky},
  journal={Advanced Functional Materials},
  volume={31},
  number={16},
  pages={2010144},
  year={2021},
  publisher={Wiley Online Library}
}

@article{du2021lead,
  title={Lead halide perovskite for efficient optoacoustic conversion and application toward high-resolution ultrasound imaging},
  author={Du, Xinyuan and Li, Jiapu and Niu, Guangda and Yuan, Jun-Hui and Xue, Kan-Hao and Xia, Mengling and Pan, Weicheng and Yang, Xiaofei and Zhu, Benpeng and Tang, Jiang},
  journal={Nature Communications},
  volume={12},
  number={1},
  pages={3348},
  year={2021},
  publisher={Nature Publishing Group UK London}
}

@article{zhang2016metal,
  title={Metal halide perovskites for energy applications},
  author={Zhang, Wei and Eperon, Giles E and Snaith, Henry J},
  journal={Nature Energy},
  volume={1},
  number={6},
  pages={1--8},
  year={2016},
  publisher={Nature Publishing Group}
}

@article{jena2019halide,
  title={Halide perovskite photovoltaics: background, status, and future prospects},
  author={Jena, Ajay Kumar and Kulkarni, Ashish and Miyasaka, Tsutomu},
  journal={Chemical Reviews},
  volume={119},
  number={5},
  pages={3036--3103},
  year={2019},
  publisher={ACS Publications}
}

@article{babayigit2016toxicity,
  title={Toxicity of organometal halide perovskite solar cells},
  author={Babayigit, Aslihan and Ethirajan, Anitha and Muller, Marc and Conings, Bert},
  journal={Nature Materials},
  volume={15},
  number={3},
  pages={247--251},
  year={2016},
  publisher={Nature Publishing Group UK London}
}

@article{lyu2017addressing,
  title={Addressing toxicity of lead: progress and applications of low-toxic metal halide perovskites and their derivatives},
  author={Lyu, Miaoqiang and Yun, Jung-Ho and Chen, Peng and Hao, Mengmeng and Wang, Lianzhou},
  journal={Advanced Energy Materials},
  volume={7},
  number={15},
  pages={1602512},
  year={2017},
  publisher={Wiley Online Library}
}

@article{slavney2017chemical,
  title={Chemical approaches to addressing the instability and toxicity of lead--halide perovskite absorbers},
  author={Slavney, Adam H and Smaha, Rebecca W and Smith, Ian C and Jaffe, Adam and Umeyama, Daiki and Karunadasa, Hemamala I},
  journal={Inorganic Chemistry},
  volume={56},
  number={1},
  pages={46--55},
  year={2017},
  publisher={ACS Publications}
}

@article{park2019intrinsic,
  title={Intrinsic instability of inorganic--organic hybrid halide perovskite materials},
  author={Park, Byung-wook and Seok, Sang Il},
  journal={Advanced Materials},
  volume={31},
  number={20},
  pages={1805337},
  year={2019},
  publisher={Wiley Online Library}
}

@article{igbari2019progress,
  title={Progress of lead-free halide double perovskites},
  author={Igbari, Femi and Wang, Zhao-Kui and Liao, Liang-Sheng},
  journal={Advanced Energy Materials},
  volume={9},
  number={12},
  pages={1803150},
  year={2019},
  publisher={Wiley Online Library}
}

@article{chu2019lead,
  title={Lead-free halide double perovskite materials: a new superstar toward green and stable optoelectronic applications},
  author={Chu, Liang and Ahmad, Waqar and Liu, Wei and Yang, Jian and Zhang, Rui and Sun, Yan and Yang, Jianping and Li, Xing’ao},
  journal={Nanomicro Lett.},
  volume={11},
  pages={1--18},
  year={2019},
  publisher={Springer}
}

@article{tang2021lead,
  title={Lead-free halide double perovskite nanocrystals for light-emitting applications: strategies for boosting efficiency and stability},
  author={Tang, Huidong and Xu, Yanqiao and Hu, Xiaobo and Hu, Qing and Chen, Ting and Jiang, Weihui and Wang, Lianjun and Jiang, Wan},
  journal={Advanced Science},
  volume={8},
  number={7},
  pages={2004118},
  year={2021},
  publisher={Wiley Online Library}
}

@article{lei2021lead,
  title={Lead-free double perovskite \ce{Cs2AgBiBr6}: fundamentals, applications, and perspectives},
  author={Lei, Hongwei and Hardy, David and Gao, Feng},
  journal={Advanced Functional Materials},
  volume={31},
  number={49},
  pages={2105898},
  year={2021},
  publisher={Wiley Online Library}
}

@article{slavney2018small,
  title={Small-band-gap halide double perovskites},
  author={Slavney, Adam H and Leppert, Linn and Saldivar Valdes, Abraham and Bartesaghi, Davide and Savenije, Tom J and Neaton, Jeffrey B and Karunadasa, Hemamala I},
  journal={Angewandte Chemie International Edition},
  volume={130},
  number={39},
  pages={12947--12952},
  year={2018},
  publisher={Wiley Online Library}
}

@article{guo2021machine,
  title={Machine learning stability and band gap of lead-free halide double perovskite materials for perovskite solar cells},
  author={Guo, Zongmei and Lin, Bin},
  journal={Sol. Energy},
  volume={228},
  pages={689--699},
  year={2021},
  publisher={Elsevier}
}

@article{khalfin2019advances,
  title={Advances in lead-free double perovskite nanocrystals, engineering band-gaps and enhancing stability through composition tunability},
  author={Khalfin, Sasha and Bekenstein, Yehonadav},
  journal={Nanoscale},
  volume={11},
  number={18},
  pages={8665--8679},
  year={2019},
  publisher={Royal Society of Chemistry}
}

@article{tailor2022dielectric,
  title={Dielectric Relaxation and Polaron Hopping in \ce{Cs2AgBiBr6} Halide Double Perovskites},
  author={Tailor, Naveen Kumar and Parikh, Nishi and Yadav, Pankaj and Satapathi, Soumitra},
  journal={The Journal of Physical Chemistry C},
  volume={126},
  number={24},
  pages={10199--10208},
  year={2022},
  publisher={ACS Publications}
}

@article{maughan2018tolerance,
  title={Tolerance factor and cooperative tilting effects in vacancy-ordered double perovskite halides},
  author={Maughan, Annalise E and Ganose, Alex M and Almaker, Mohammed A and Scanlon, David O and Neilson, James R},
  journal={Chemistry of Materials},
  volume={30},
  number={11},
  pages={3909--3919},
  year={2018},
  publisher={ACS Publications}
}

@article{li2017high,
  title={High-pressure band-gap engineering in lead-free \ce{Cs2AgBiBr6} double perovskite},
  author={Li, Qian and Wang, Yonggang and Pan, Weicheng and Yang, Wenge and Zou, Bo and Tang, Jiang and Quan, Zewei},
  journal={Angewandte Chemie International Edition},
  volume={56},
  number={50},
  pages={15969--15973},
  year={2017},
  publisher={Wiley Online Library}
}

@article{zhang2019tuning,
  title={Tuning emission and electron--phonon coupling in lead-free halide double perovskite \ce{Cs2AgBiCl6} under pressure},
  author={Zhang, Long and Fang, Yuanyuan and Sui, Laizhi and Yan, Jiejuan and Wang, Kai and Yuan, Kaijun and Mao, Wendy L and Zou, Bo},
  journal={ACS Energy Letters},
  volume={4},
  number={12},
  pages={2975--2982},
  year={2019},
  publisher={ACS Publications}
}

@article{fabini2020underappreciated,
  title={The underappreciated lone pair in halide perovskites underpins their unusual properties},
  author={Fabini, Douglas H and Seshadri, Ram and Kanatzidis, Mercouri G},
  journal={MRS Bulletin},
  volume={45},
  number={6},
  pages={467--477},
  year={2020},
  publisher={Cambridge University Press}
}

@article{wiktor2023quantifying,
  title = {Quantifying Dynamic Tilting in Halide Perovskites: Chemical Trends and Local Correlations},
  author = {Wiktor, Julia and Fransson, Erik and Kubicki, Dominik and Erhart, Paul},
  year = 2023,
  month = {4},
  journal = {Chemistry of Materials},
  volume = {35},
  pages = {6737},
  arxiv = {2304.07402},
  doi = {10.1021/acs.chemmater.3c00933},
  zenodo = {7889313},
}

@article{meyer2018lead,
  title={Lead-free halide double perovskites: a review of the structural, optical, and stability properties as well as their viability to replace lead halide perovskites},
  author={Meyer, Edson and Mutukwa, Dorcas and Zingwe, Nyengerai and Taziwa, Raymond},
  journal={Metals},
  volume={8},
  number={9},
  pages={667},
  year={2018},
  publisher={MDPI}
}

@article{volonakis2016lead,
  title={Lead-free halide double perovskites via heterovalent substitution of noble metals},
  author={Volonakis, George and Filip, Marina R and Haghighirad, Amir Abbas and Sakai, Nobuya and Wenger, Bernard and Snaith, Henry J and Giustino, Feliciano},
  journal={The Journal of Physical Chemistry Letters},
  volume={7},
  number={7},
  pages={1254--1259},
  year={2016},
  publisher={ACS Publications}
}

@article{caicedo2024disentangling,
  title={Disentangling the effects of structure and lone-pair electrons in the lattice dynamics of halide perovskites},
  author={Caicedo-D{\'a}vila, Sebasti{\'a}n and Cohen, Adi and Motti, Silvia G and Isobe, Masahiko and McCall, Kyle M and Grumet, Manuel and Kovalenko, Maksym V and Yaffe, Omer and Herz, Laura M and Fabini, Douglas H and others},
  journal={Nature Communications},
  volume={15},
  number={1},
  pages={4184},
  year={2024},
  publisher={Nature Publishing Group UK London}
}

@article{ovito,
  title={{Visualization and analysis of atomistic simulation data with OVITO--the Open Visualization Tool}},
  author={Stukowski, Alexander},
  journal={Modelling and Simulation in Materials Science and Engineering},
  volume={18},
  number={1},
  pages={015012},
  year={2009},
  publisher={IOP Publishing}
}

@article{larsen2016robust,
  title={Robust structural identification via polyhedral template matching},
  author={Larsen, Peter Mahler and Schmidt, S{\o}ren and Schi{\o}tz, Jakob},
  journal={Modelling and Simulation in Materials Science and Engineering},
  volume={24},
  number={5},
  pages={055007},
  year={2016},
  publisher={IOP Publishing}
}

@article{scipy,
  title={{SciPy} 1.0: fundamental algorithms for scientific computing in {Python}},
  author={Virtanen, Pauli and Gommers, Ralf and Oliphant, Travis E and Haberland, Matt and Reddy, Tyler and Cournapeau, David and Burovski, Evgeni and Peterson, Pearu and Weckesser, Warren and Bright, Jonathan and others},
  journal={Nature Methods},
  volume={17},
  number={3},
  pages={261--272},
  year={2020},
  publisher={Nature Publishing Group}
}

@article{fan2021neuroevolution,
  title={Neuroevolution machine learning potentials: Combining high accuracy and low cost in atomistic simulations and application to heat transport},
  author={Fan, Zheyong and Zeng, Zezhu and Zhang, Cunzhi and Wang, Yanzhou and Song, Keke and Dong, Haikuan and Chen, Yue and Ala-Nissila, Tapio},
  journal={Physical Review B},
  volume={104},
  number={10},
  pages={104309},
  year={2021},
  publisher={APS}
}

@article{fan2022gpumd,
  title={GPUMD: A package for constructing accurate machine-learned potentials and performing highly efficient atomistic simulations},
  author={Fan, Zheyong and Wang, Yanzhou and Ying, Penghua and Song, Keke and Wang, Junjie and Wang, Yong and Zeng, Zezhu and Xu, Ke and Lindgren, Eric and Rahm, J Magnus and others},
  journal={The Journal of Chemical Physics},
  volume={157},
  number={11},
  year={2022},
  publisher={AIP Publishing}
}

@article{bernetti2020pressure,
  title={Pressure control using stochastic cell rescaling},
  author={Bernetti, Mattia and Bussi, Giovanni},
  journal={The Journal of Chemical Physics},
  volume={153},
  number={11},
  year={2020},
  publisher={AIP Publishing}
}

@article{glazer1972classification,
  title={The classification of tilted octahedra in perovskites},
  author={Glazer, Anthony M},
  journal={Acta Crystallographica Section B: Structural Crystallography and Crystal Chemistry},
  volume={28},
  number={11},
  pages={3384--3392},
  year={1972},
  publisher={International Union of Crystallography}
}

@article{howard1998group,
  title={Group-theoretical analysis of octahedral tilting in perovskites},
  author={Howard, Christopher J and Stokes, Harold T},
  journal={Acta Crystallographica Section B: Structural Science},
  volume={54},
  number={6},
  pages={782--789},
  year={1998},
  publisher={International Union of Crystallography}
}

@article{xu2025mega,
    title={{GPUMD 4.0: A high-performance molecular dynamics package for versatile materials simulations with machine-learned potentials}},
    author={K. Xu and H. Bu and S. Pan and E. Lindgren and Y. Wu and Y. Wang and J. Liu and K. Song and B. Xu and Y. Li and T. Hainer and L. Svensson and J. Wiktor and R. Zhao and H. Huang and C. Qian and S. Zhang and Z. Zeng and B. Zhang and B. Tang and Y. Xiao and Z. Yan and J. Shi and Z. Liang and J. Wang and T. Liang and S. Cao and Y. Wang and P. Ying and N. Xu and C. Chen and Y. Zhang and Z. Chen and X. Wu and W. Jiang and E. Berger and Y. Li and S. Chen and A. J. Gabourie and H. Dong and S. Xiong and N. Wei and Y. Chen and J. Xu and F. Ding and Z. Sun and T. Ala-Nissila and A. Harju and J. Zheng and P. Guan and P. Erhart and J. Sun and W. Ouyang and Y. Su and Z. Fan},
    journal={Materials Genome Engineering Advances},
    volume={3},
    pages={e70028},
    doi={doi: 10.1002/mgea.70028},
    year={2025}
}

@article{song2024general,
author = {Song, Keke and Zhao, Rui and Liu, Jiahui and Wang, Yanzhou and Lindgren, Eric and Wang, Yong and Chen, Shunda and Xu, Ke and Liang, Ting and Ying, Penghua and Xu, Nan and Zhao, Zhiqiang and Shi, Jiuyang and Wang, Junjie and Lyu, Shuang and Zeng, Zezhu and Liang, Shirong and Dong, Haikuan and Sun, Ligang and Chen, Yue and Zhang, Zhuhua and Guo, Wanlin and Qian, Ping and Sun, Jian and Erhart, Paul and Ala-Nissila, Tapio and Su, Yanjing and Fan, Zheyong},
title = {General-purpose machine-learned potential for 16 elemental metals and their alloys},
journal = {Nature Communications},
volume = {15},
number = {1},
pages = {10208},
year = {2024},
doi = {10.1038/s41467-024-54554-x}
}

@article{FraWikErh23,
  title = {Phase transitions in inorganic halide perovskites from machine-learned potentials},
  author = {Fransson, Erik and Wiktor, Julia and Erhart, Paul},
  year = 2023,
  month = {7},
  journal = {The Journal of Physical Chemistry C},
  volume = {127},
  pages = {13773},
  arxiv = {2301.03497},
  doi = {10.1021/acs.jpcc.3c01542},
}

\end{document}